\documentclass{optica-article}

\journal{opticajournal} 
\articletype{Research Article}

\usepackage{graphicx}
\usepackage{subcaption}

\usepackage{lineno}

\begin{document}

\title{Integrated diode lasers for the generation of sub-GHz repetition rate frequency combs}

\author{A. Memon,\authormark{1, *}, A. van Rees,\authormark{1, 2}, J. Mak, \authormark{1}, Y. Fan,\authormark{1}, P.J.M van der Slot,\authormark{1}, H.M.J. Bastiaens,\authormark{1}  and K.-J. Boller\authormark{1}}

\address{\authormark{1}Laser Physics and Nonlinear Optics, Department for Science and Technology, University of Twente, Enschede, The Netherlands
\\
\authormark{2}Currently with Chilas B.V, Enschede, The Netherlands}

\email{\authormark{*}a.memon@utwente.nl} 


\begin{abstract*} 
We demonstrate absorber-free passive and hybrid mode-locking at sub-GHz repetition rates using a hybrid integrated extended cavity diode laser around 1550 nm. The laser is based on InP as gain medium and a long Si$_3$N$_4$ feedback circuit, with three highly frequency selective microring resonators. The feedback resonators not only increases the cavity length up to 0.6 m to achieve sub-GHz repetition rates but also serve as a dispersive narrowband mirror for sharp spectral filtering, which enables Fourier domain mode-locking. We observe passive mode-locking with repetition rates below 500 MHz, with $\approx$ 15 comb lines at around 0.2 mW total power. To stabilize the repetition rate, hybrid mode-locking is demonstrated by weak RF modulation of the diode current at frequencies around 500 MHz. The RF injection reduces the Lorentzian linewidth component from 8.9 kHz to a detection limited value around 300 mHz. To measure the locking range of the repetition rate, the injected RF frequency is tuned with regard to the passive mode-locking frequency and the injected RF power is varied. The locking range increases approximately as a square-root function of the injected RF power. At 1 mW  injection a wide locking range of about 80 MHz is obtained. We observe the laser maintaining stable mode-locking also when  the DC diode pump current is increased from 40 mA to 190 mA, provided that the cavity length is maintained constant with thermo-refractive tuning. 

\end{abstract*}

\section{Introduction}
Optical frequency combs (OFCs) have gained considerable interest in fundamental science and in  applications such as metrology \cite{OFC}. Of particular interest are dense frequency combs, i.e., with small mode spacing through low, sub-GHz, repetition rates, for instance, to resolve the typically GHz-wide absorption lines of atmospheric trace gases \cite{atmosphericGasPaper}. Further potential applications of low repetition rate frequency combs could be enabled by miniaturization through photonic integration with electrically pumped semiconductor optical amplifiers. Examples are upscaling of quantum optical systems \cite{QuantumKues, QuantumMIT, QuantumTaballione, QuantumHatam} or compact LIDAR sensors \cite{SwannLidar}. In all instances, maintaining a high level of frequency stability is essential.\\


Conventionally, providing low sub-GHz repetition rates has been based on bulk lasers with long (meter-sized) resonators, typically Ti:Sapphire lasers and rare-earth doped fiber lasers. A central property of such gain media that enables stable mode-locking in the form of pulses is that the gain lifetime is much longer than the cavity roundtrip time (tens to hundreds of microseconds as compared to typically around ten nanoseconds, respectively). Nevertheless, such lasers exhibit complexity in operation due to the need for optical pumping, while physically long bulk cavities make these systems susceptible to mechanical and acoustic perturbations. These features, including size and weight, limit bulk lasers predominately to lab usage. \\


To address the named drawbacks, there has been a push to develop frequency combs based on integrated photonics, which offers compact, chip-based solutions that can be scaled toward more complex systems. An example is generating Kerr-combs in microring resonators \cite{kippenbegGeneralKerr, kerrcombs}. These combs can provide extremely wide spectra, however, the repetition rates are high, typically above 50 GHz, and the conversion efficiency is typically only at the percent-level \cite{HighRepRateLowConversion}. Monolithic mode-locked diode lasers provide the unique advantage of direct electrical pumping, eliminating the need for optical pumping and thus provides higher optical power and overall efficiencies \cite{MLLHighOutpower}. Electrical pumping and monolithic integration of the laser cavity strongly decreases the susceptibility to acoustic perturbations and simplifies the over all system, including thermal management. However, monolithic mode-locked diode lasers share a drawback with Kerr frequency combs in that they also exhibit high, GHz repetition rates, due to their short cavity length. A fundamental limitation of monolithic diode lasers is the broader intrinsic linewidth of the comb frequencies \cite{MonolithicMHzLW, monolithic398khz}, which is due to the short cavity length as well, in combination with a relatively high intrinsic waveguide loss in diode laser amplifiers. \\ 

An interesting solution for lowering the repetition rate and also the intrinsic linewidth of diode laser is to increase the cavity length by heterogeneous integration \cite{HeterogenousGhent} and hybrid integration \cite{LPNOHybrid}. These techniques involve the use of extended cavities such as made of low-loss Si$_3$N$_4$ feedback circuits and integrated saturable absorbers to generate mode-locked pulses. Although such feedback circuits can provide long roundtrip length in a chip-size format, there is a fundamental stability problem. Lowering the repetition rates into the sub-GHz range is prone to introduce chaotic fluctuations \cite{YoonChaoticfluct, ChaoticFlucTheory} or harmonic mode-locking \cite{HarmonicMLL} with several pulses per roundtrip, the latter increasing the repetition rate again. The reasons are the short gain lifetimes and recovery times that are typically around 1 ns in semiconductor optical amplifiers (SOA)\cite{lifetimeRP}. If the roundtrip time of the pulses in the laser resonator becomes longer than the upper state lifetime, amplified spontaneous emission between pulses destabilizes mode-locking \cite{ASEDestabliseML}. This limitation makes it difficult to achieve the desired sub-GHz repetition rates with diode lasers using extended cavities, whether using bulk optics or integrated photonics.\\

To open the path toward sub-GHz repetition rate mode-locking with integrated diode lasers, here we explore the alternative approach of Fourier domain mode-locking (FDML) \cite{FDMLHuber}. With FDML, in order to avoid amplitude modulation, the laser cavity does not contain a saturable absorber. Instead, a comb spectrum is generated through frequency modulation synchronous with the roundtrip time rate, with the output remaining quasi-continuous. Thereby, stimulated emission continuously saturates the gain, which suppresses amplified spontaneous emission independent of the cavity roundtrip time. The property of FDML being intrinsically absorber-free is also of practical relevance. For the telecom wavelength range, saturable absorbers integrated with semiconductor amplifiers are available within multi-project wafer (MPW) runs, however, other wavelength ranges require dedicated fabrication involving much higher costs. These fundamental differences and practical aspects makes FDML more universal, as sub-GHz generation can presumably be obtained in many wavelength ranges due to wide availability of semiconductor amplifiers.\\

In recent experiments, FDML based on diode lasers has been successfully demonstrated by using spectral feedback filtering \cite{Ibrahimi_Thales, Yuan_chinesepaper}. We note that spectral filtering, such as with an intra-cavity grating, is required also with standard (pulsed) mode-locking to restrict the destabilizing influence of intra-cavity dispersion. In contrast, to obtain stable FDML, sharper spectral filtering is required. The qualitative explanation is that the filter-induced bandwidth reduction and the associated, steep filter dispersion counteract ultrashort pulse formation and imposes quasi-continuous oscillation. To more detail \cite{Mariangela} detuned loading introduced by the filter \cite{vahala_Detunedloading} enables amplification of relaxation oscillation (RO), if the RO frequency is in resonance with the mode spacing frequency. The RO then generates multiple sidebands in the laser cavity modes, such that single-frequency oscillation is suppressed.\\

The lowest repetition rates reported with this approach with diode lasers so far are 255 MHz \cite{Yuan_chinesepaper} and 360 MHz \cite{Ibrahimi_Thales}. However, for the former, bulk optics was used for cavity length extension and spectral filtering (a lens-coupled Bragg fiber), which is not suitable for chip-integration. In the other case (360 MHz), the laser was fully integrated with a SiN waveguide delay line for cavity length extension while a Bragg waveguide served for spectral feedback filtering. The observed average repetition rate was matching the extended cavity length. However, the authors also noted signs of dynamic instability due to independent locking of subgroups of modes, which we address to an insufficiently narrow Bragg filtering. In contrast, stable FDML has been observed in hybrid integrated lasers enabled by the sharper spectral filtering via microring resonators \cite{Jesse, Yvanklaver}. The repetition rates of these lasers were, however, about an order of magnitude higher than in \cite{Yuan_chinesepaper} and \cite{Ibrahimi_Thales}, i.e., 5 GHz and 2.5 GHz, respectively.\\

Here, we demonstrate stable Fourier domain mode-locking in a chip integrated, long cavity diode laser for the first time at sub-GHz ($\approx$ 500 MHz) repetition rates. The extended cavity diode laser is based on hybrid integration of an InP semiconductor amplifier chip with a low-loss Si$_3$N$_4$ feedback circuit. The low repetition rate is achieved by double-passing the intra-cavity light through three sequential microring resonators (MRRs) which extends the laser cavity roundtrip length beyond half a meter via multiple passes through the MRRs. We demonstrate passive mode-locking tunable around 500 MHz repetiton rate via the laser cavity parameters. Hybrid mode-locking is achieved by driving the diode laser with an additional RF current at or around the passive repetiton rate. The injected RF current stabilizes repetition rate which is observed as linewidth reduction in the laser's RF spectrum. Within a locking range of up to 80 MHz, the laser's repetition rate can be tuned with the applied RF frequency.

\section{Experimental Results}

\subsection{Hybrid Integrated Laser}


\begin{figure}[ht!]
\centering\includegraphics[width=12cm]{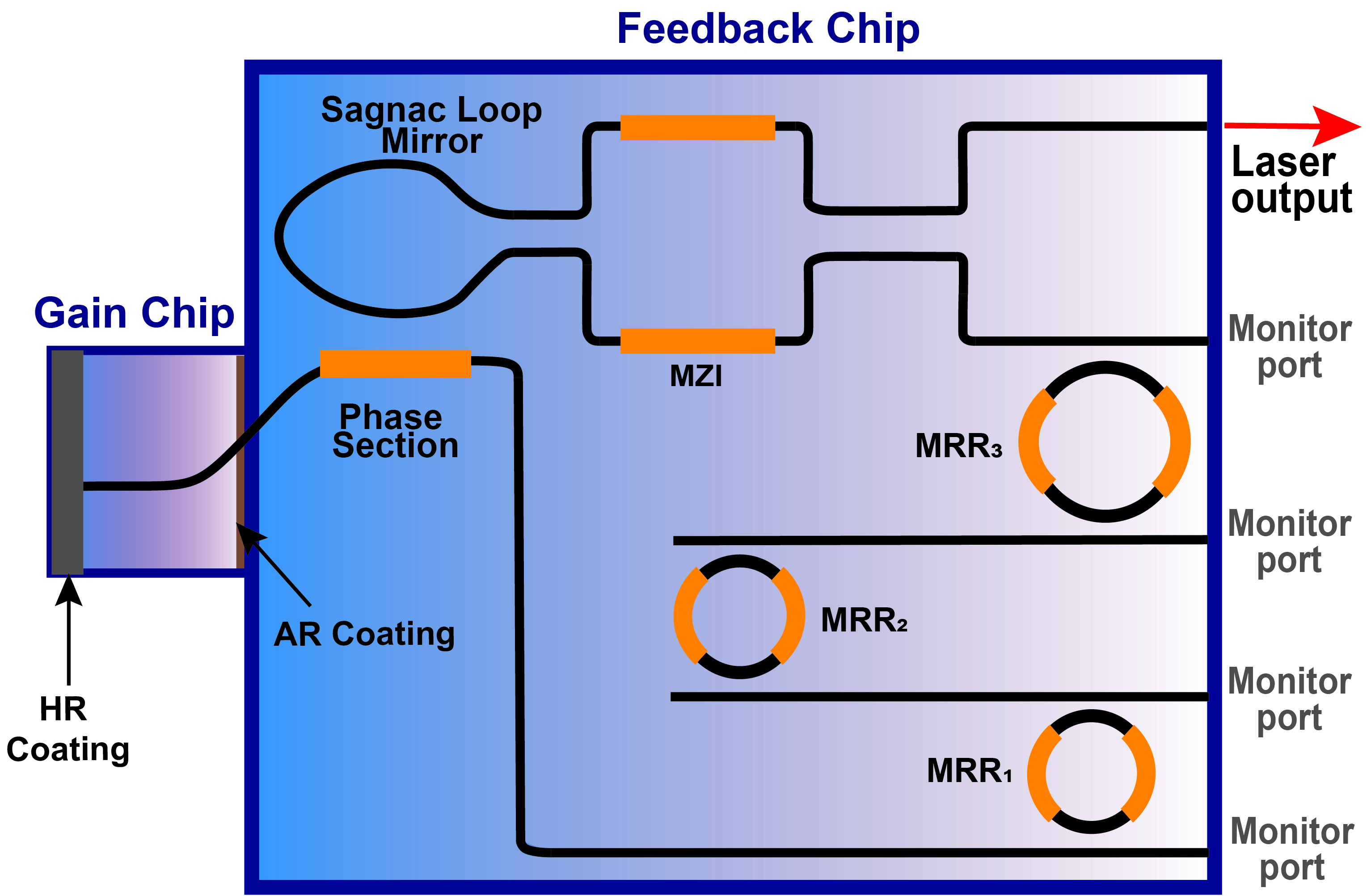}
\caption{Schematic diagram of hybrid integrated mode-locked laser (not to scale). The laser comprises of two parts. The gain chip is highly reflective (HR) coated on the left-hand side. The other end is anti-reflective (AR) coated to reduce undesired back reflection from the interface between the two chips. Residual back reflections from the InP and Si$_3$N$_4$ facets are diverted out of the circuit through tilting the InP waveguide at 9 $^{\circ}$ relative to the facet normal. The feedback circuit is based on Si$_3$N$_4$, using three micro ring resonators (MRR$_1$, MRR$_2$ and MRR$_3$) as Vernier filter. A phase section is used to tune the frequency of the longitudinal laser cavity modes relative to the Vernier filter frequency. The feedback circuit is completed with a Sagnac loop mirror, the reflectivity of which can be controlled by a Mach–Zehnder interferometer (MZI). The MZI, MRRs and phase section are equipped with thermoelectric heaters (shown in yellow). All output ports from the chip are connected to optical fibers.}

\label{fig:Laserschemetic}
\end{figure}

The laser used in the experiments is shown schematically in Fig. \ref{fig:Laserschemetic}. The circuit design is comparable to a previous version \cite{Fan40Hz}, except for a somewhat longer optical cavity length of 60 cm. The laser comprises two parts. The first part is an InP gain chip, and the second part is a feedback circuit which is made of low-loss Si$_3$N$_4$ waveguides. The two chips are hybrid integrated by chemical bonding after edge coupling. \\ 

The InP gain chip is fabricated by the Fraunhofer Heinrich Hertz Institute (HHI, Berlin), carrying a double-pass semiconductor optical amplifier (SOA), with a 100 nm wide gain spectrum at a center wavelength around 1520 nm. The back side of the gain chip is highly-reflective (HR) coated. To minimize back-reflections into the amplifier, the front side of the gain chip is anti-reflective (AR) coated. Undesired effects from residual facet reflection are further reduced through tilting the InP waveguide at 9 $^{\circ}$ relative to the facet normal. In order to optimize matching the feedback waveguide mode to that of the gain waveguide, we use tapering and a Fresnel-refraction matched tilt of 20 $^{\circ}$ \cite{LionixSi3N4explaination}. \\

The feedback circuit is based on asymmetric double-stripe (ADS) Si$_3$N$_4$ waveguides, fabricated by LioniX International \cite{LionixSi3N4explaination}. The circuit comprises a Vernier spectral filter consisting of three microring resonators in (MRRs) in series with different radii (R$_1$ = 150 $\mu$m, R$_2$ = 153 $\mu$m and R$_3$ = 1500 $\mu$m) and a power coupling coefficient of $\kappa$ = 10\%. With such coupling the light performs approximately \(\frac{1-\kappa}{\kappa}\)= 9 roundtrips \cite{KappaIncreasingLength} in each MRR, when fully at resonance. Completing the laser cavity with a Sagnac waveguide mirror lets the light pass twice per rountrip through the three MRRs. This extends the overall optical cavity roundtrip length to approximately 0.6 m when setting the MRR transmission resonances to the same wavelength via thermo-refractive tuning. The MRRs are also responsible for introducing a steep dispersion in the laser cavity which is essential for comb generation.\\

An adjustable part of the intracavity light is coupled out at the Sagnac loop mirror with a tunable Mach-Zehnder interferometer (MZI). The optical length of the cavity can be fine tuned with a phase section, which tunes the cavity mode frequencies with regard to the Vernier filter transmission. For thermo-electric tuning of the MZI, the MRRs and the phase section, the feedback chip is equipped with thin-film heaters. For stable operation the laser is  fully packaged. This includes electric wire bonding for the electrical connections, mode-matched, polarization maintaining  fibers  bonded to the laser output waveguide, and to the other ports and mounting the chips on a Peltier cooler for temperature control. 


\subsection{Experimental setup}

\begin{figure}[ht!]
\centering\includegraphics[width=13cm]{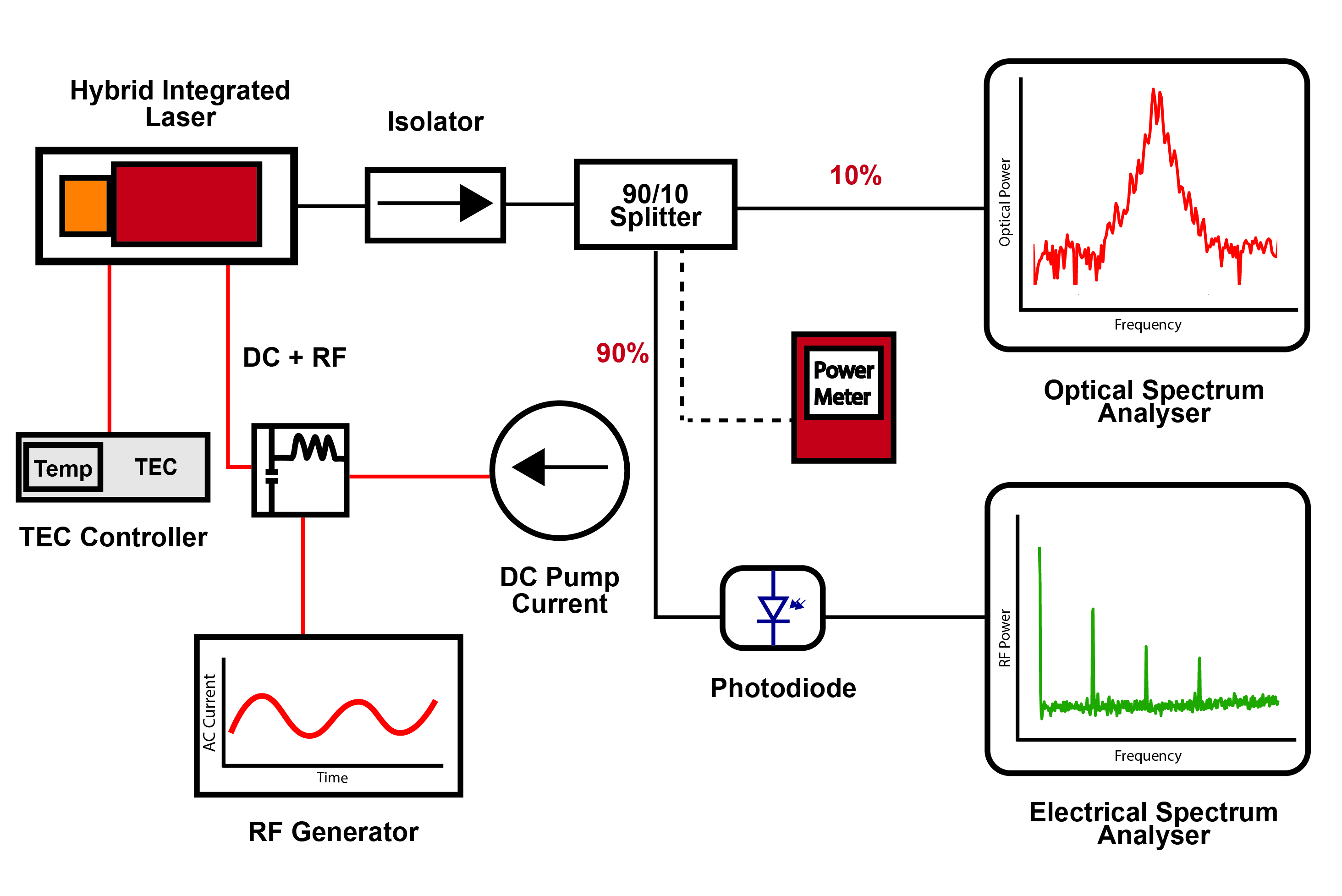}
\caption{Schematic view of the setup for realizing mode-locking. The figure shows the hybrid laser, to which a DC and AC current is applied, combined via a bias-T. The laser is also connected to a temperature controller. The laser output is sent via a fiber isolator and 90/10 splitter to an optical spectrum analyzer (OSA) and a power meter, or to a fast photodiode followed by an electrical spectrum analyzer (ESA). }
\label{fig:Setup}
\end{figure}

The experimental setup for mode-locking experiments is shown in Fig. \ref{fig:Setup}. The hybrid laser is pumped with a DC current source (LDX-3207 Precision Current Source). For stabilizing the temperature, a temperature controller is used (LDT-5910). The output from the laser is characterized behind an isolator (Thorlabs IO-G-1550-APC) using an optical spectrum analyzer (OSA, Finisar WaveAnalyzer 1500S, highest resolution 150 MHz). To observe the output with higher resolution via measuring the underlying mode beating frequencies, a smaller fraction of the output light is sent via a fiber splitter (Thorlabs  1550 ± 100 nm, 90:10) to a fast photodiode (Discovery Semiconductors, 20 GHz bandwidth) connected to an electrical spectrum analyzer (ESA, Keysight N9000B CXA), alternately to power meter (Thorlabs, PM100D). For hybrid mode-locking an additional AC current is applied to the gain section using an RF signal generator (69147A - Anritsu Sweep Generator, frequency range 10 MHz - 20 GHz). A bias-T (Mini-Circuits Bias-TEE, ZFBT-4R2GW, 0.1 MHz - 4200 MHz) is used to add the AC current for gain modulation.\\


\subsection{Passive Mode-Locking}
\label{PMLtext}

In the following, we describe sub-GHz generation of frequency combs by passive mode-locking, which is achieved by solely applying a direct current (DC) to the gain section.  The threshold pump current was measured to be about 27 mA. The experiment was carried out by first tuning the laser to single-frequency oscillation, using a pump current of 50 mA which typically generates an output power of 1 mW. Single-frequency output is obtained by tuning the laser output to near the center of the gain spectrum, using coarse adjustments of the heater currents for MRR$_1$ and MRR$_2$ similar to what is described in \cite{Fan40Hz}. Next, the heater for MRR$_3$ is adjusted as well until the laser output power is maximized simultaneously with observing a single output frequency with the OSA. The final step of preparation is tuning the phase section to maximize the power of the single-frequency output. \\

    

\begin{figure}[ht!]
    \centering
    \begin{subfigure}{0.49\textwidth}
        \includegraphics[width=\textwidth]{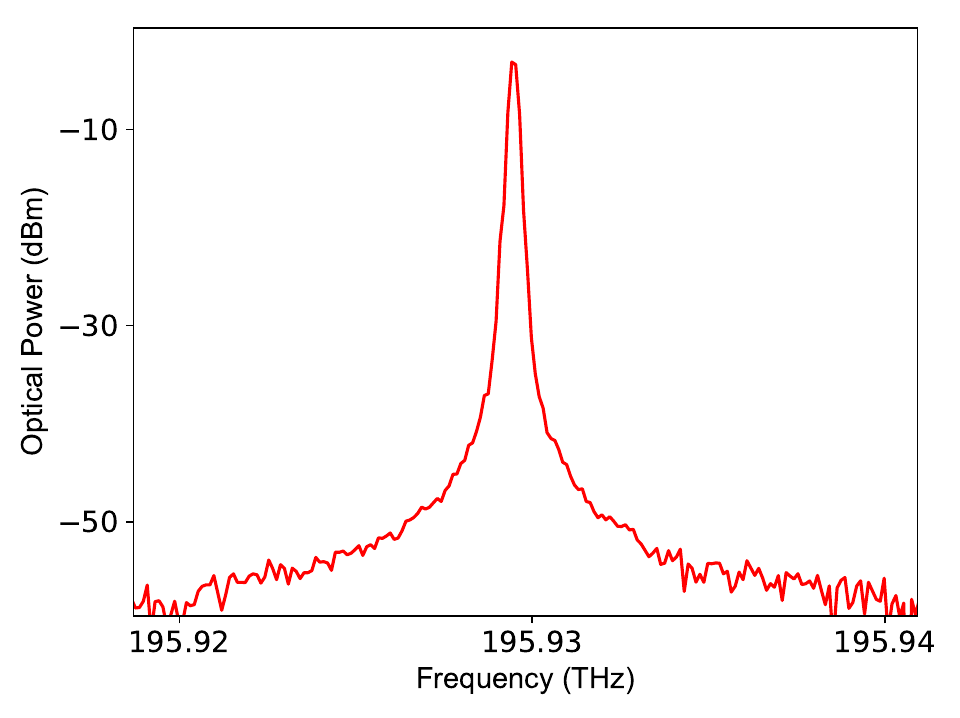}
        \end{subfigure}
    \begin{subfigure}{0.49\textwidth}
        \includegraphics[width=\textwidth]{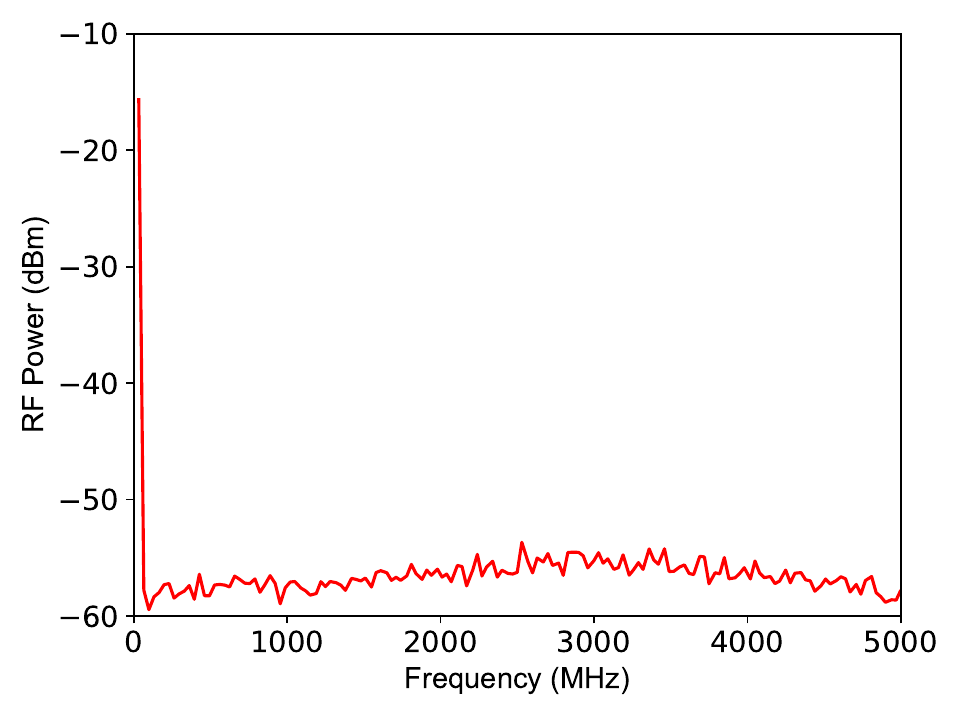}
   \end{subfigure}
    \caption{(a) Optical spectrum recorded with the optical spectrum analyzer (OSA, resolution 150 MHz) when the laser is operating in a single mode. (b) The corresponding RF spectrum recorded on the electrical spectrum analyzer (ESA, resolution 3 MHz). The ESA spectrum does not display any mode-beating related peaks, confirming single frequency operation.}
  \label{fig:PML}
\end{figure}

Figure \ref{fig:PML}a shows a typical optical power spectrum recorded with a resolution of 150 MHz, indicating single-frequency oscillation within the resolution of the OSA. Figure \ref{fig:PML}b shows the simultaneously recorded RF spectrum. The spectrum only shows a flat spectrum close to the noise floor of the ESA. The absence of any beat notes confirms single-frequency emission. \\

\begin{figure}[ht!]
    \centering
    \begin{subfigure}{0.49\textwidth}
        \includegraphics[width=\textwidth]{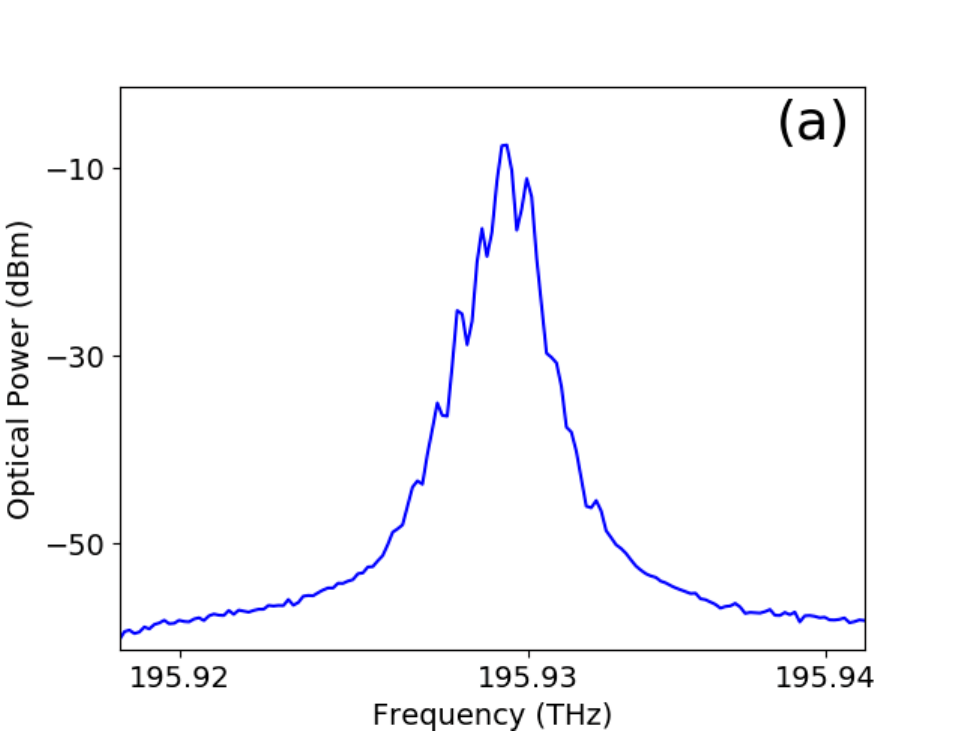}
            \end{subfigure}
    \begin{subfigure}{0.49\textwidth}
        \includegraphics[width=\textwidth]{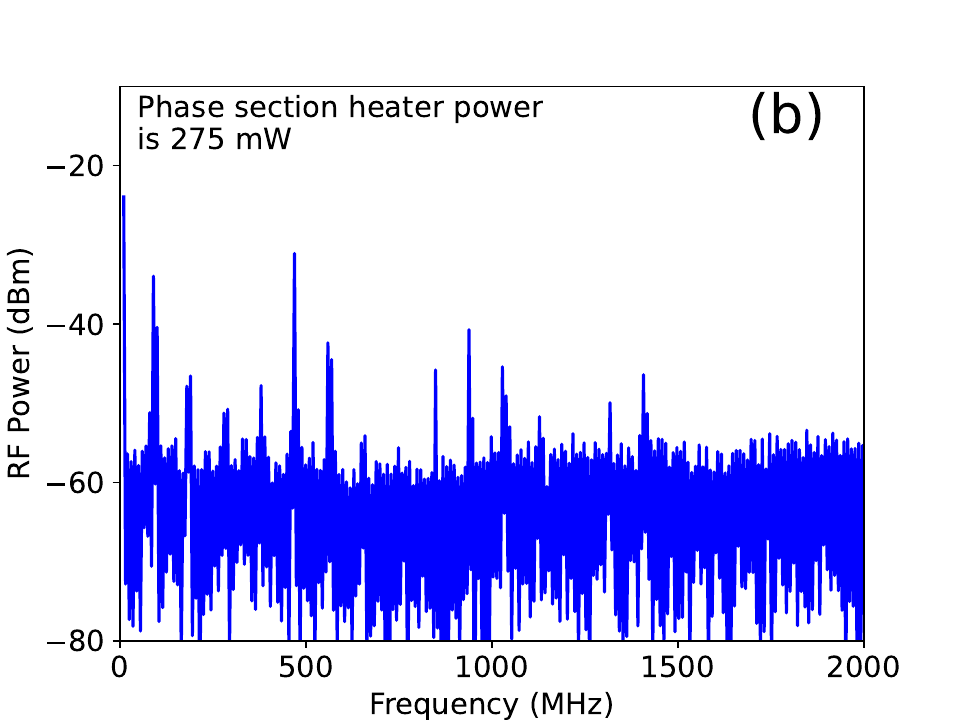}
          \end{subfigure}
    \begin{subfigure}{0.49\textwidth}
        \includegraphics[width=\textwidth]{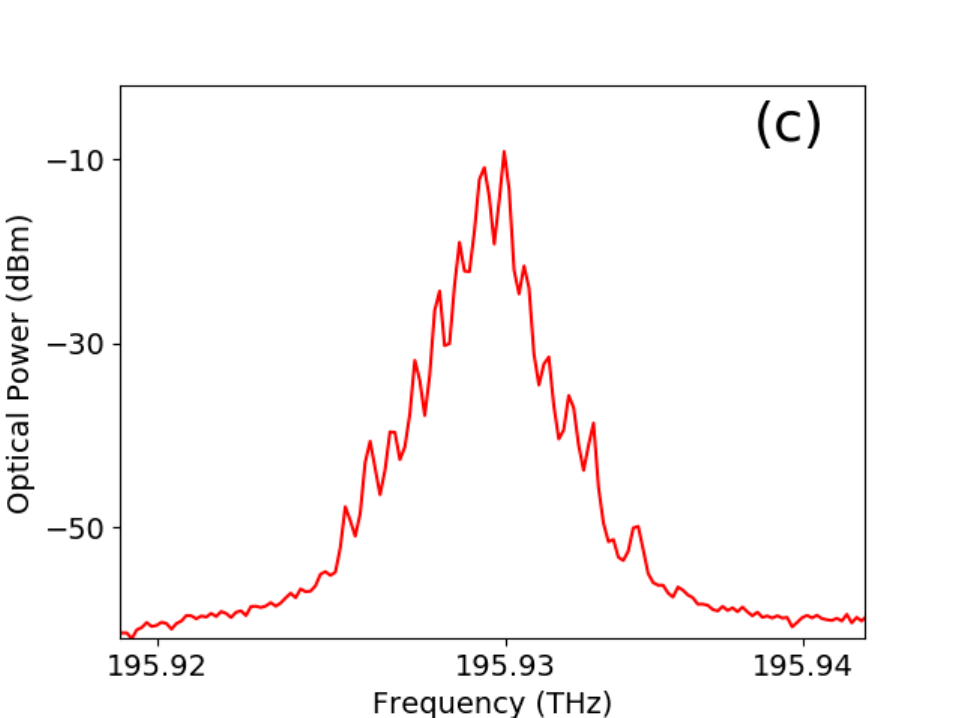}
            \end{subfigure}
    \begin{subfigure}{0.49\textwidth}
        \includegraphics[width=\textwidth]{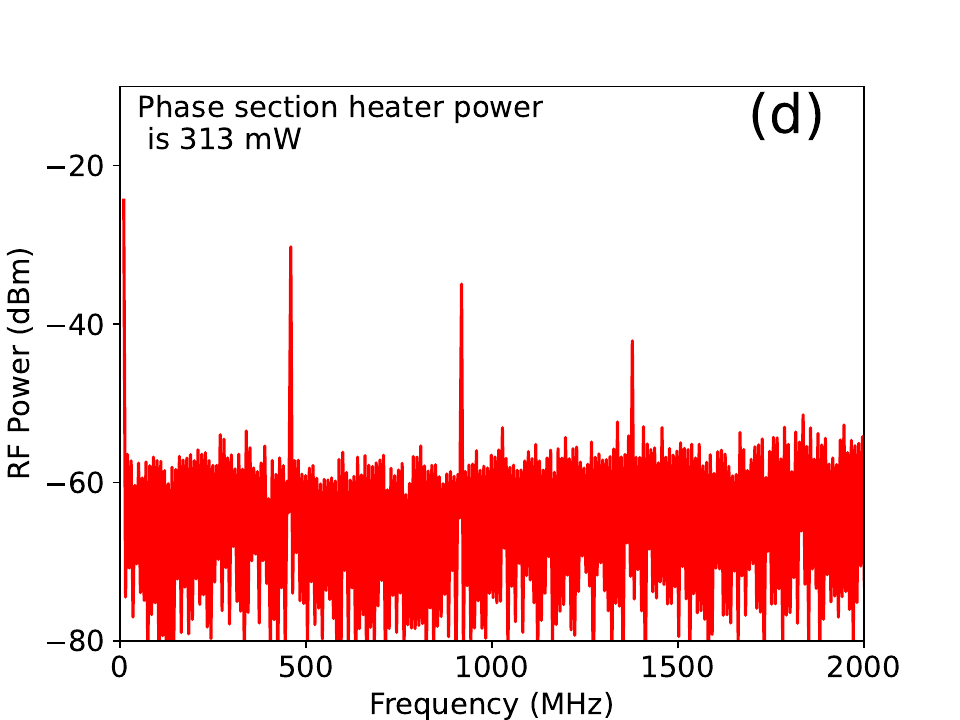}
        \end{subfigure}
    \caption{(a) Optical spectrum recorded on OSA when the laser is operating in multimode. (b) The corresponding RF spectrum recorded with the ESA when the laser operates in multimode. (c) Optical spectrum recorded with the OSA, when the rings and phase sections are tuned and laser is operating as a mode-locked laser. (d) The corresponding RF spectrum at the ESA when the laser operates as a mode-locked laser.}
    \label{fig:Multimode}
\end{figure}

In order to generate a frequency comb by passive mode-locking, the phase section is tuned toward longer optical cavity length as theoretically expected \cite{GioanniniHighPumpCurrent} via increasing the heater power. As the intermediate step toward frequency comb generation, we first observe multimode oscillation. Figure \ref{fig:Multimode}a presents an example of the output spectrum from the OSA when the laser operates in multiple modes. However, due to insufficient resolution of the OSA, distinguishing multimode operation (with non-uniform frequency spacing due to intracavity dispersion) from mode-locked operation with strictly equidistant spacing is not possible. For inspecting the output with much higher resolution at the kHz-level, we inspect the laser's RF spectrum as shown in Fig. \ref{fig:Multimode}. The RF spectrum displays the presence and spacings of multiple optical frequencies via their RF beating frequencies. It can be seen that the RF spectrum comprises a larger variety of beat frequencies with different spacings. We address this to the presence of chaotic dynamics as described in \cite{Mariangela}. \\

To generate an optical frequency comb, we continue to increase the heater current of the phase section while monitoring both the spectra at the OSA and ESA. Figure \ref{fig:Multimode}c shows the optical spectrum after increasing the phase section heating by 38 mW. Compared to Fig. \ref{fig:Multimode}a it can be seen that the number and contrast of the side peaks has increased. A clear change occurs in the RF spectrum as depicted in Fig. \ref{fig:Multimode}d. The spectrum now consists of a set of narrowband peaks with only a single fundamental RF frequency around 451 MHz and integer multiples of it. The width of the fundamental peak lies in the kilohertz range (see next section), thus, an optical frequency comb is generated. We note that the measured repetiton rate corresponds to a laser cavity roundtrip time of approximately 2.2 ns, which agrees well with the estimation of the laser's optical  roundtrip time of 0.6 m (2.0 ns). With these values, the roundtrip time is a factor of 2.2 bigger than the approximate 1-ns gain lifetime. For pulsed mode-locking using saturable absorbers, the lowest stable repetition rate  would be higher, at around 1 GHz. We note that the mode-locking observed here resembles Fourier domain mode-locking as previously observed in \cite{Mariangela} and \cite{Jesse}, though with up to an order of magnitude lower repetition rate.\\

For tuning of the comb line spacing such as might be required, e.g., for dual comb measurements \cite{DualCombHetrodyne} we recall that the comb spacing varies with the optical roundtrip length of the laser cavity, which should be tunable via the microring resonators. A slight detuning of one of the ring resonances from each other should lower the number of MRR roundtrips performed, thereby shortening the optical cavity length thus increasing the spacing of the comb lines. A potential problem that might inhibit such repetition rate tuning is that it also changes the amount of detuned loading on which the mode-locking dynamics is based. It is thus not clear beforehand in how far the mode-locking frequency can be tuned. \\

\begin{figure}[ht!]
    \centering
    \begin{subfigure}{0.49\textwidth}
    \centering
    \includegraphics[width=1\linewidth]{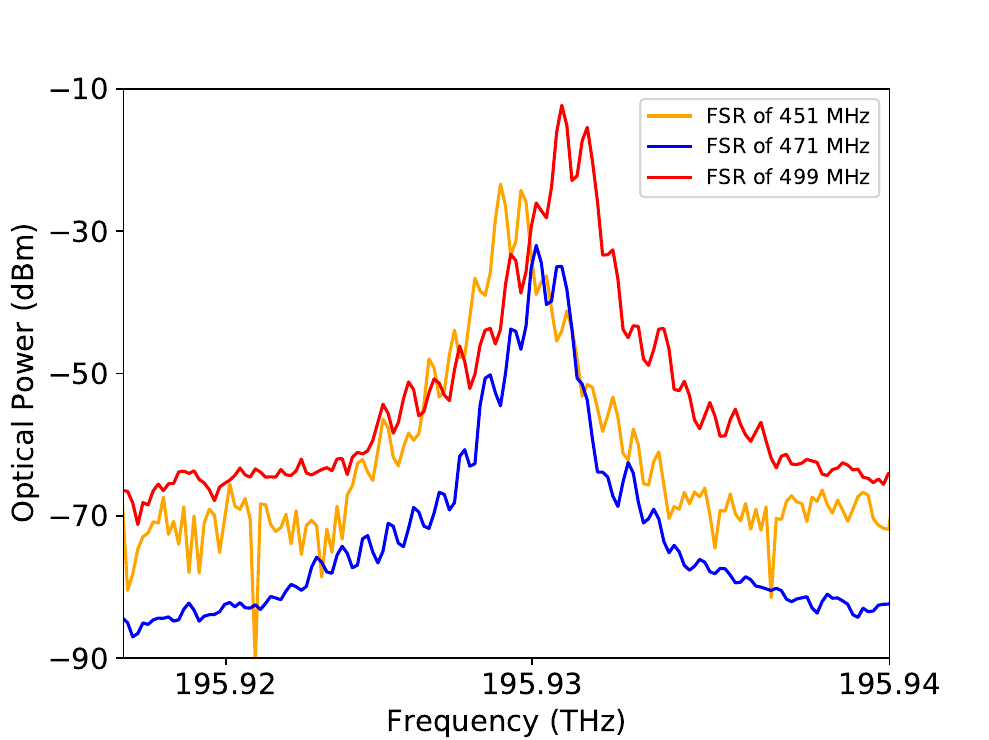}
    \end{subfigure}
     \begin{subfigure}{0.49\textwidth}
    \includegraphics[width=1\linewidth]{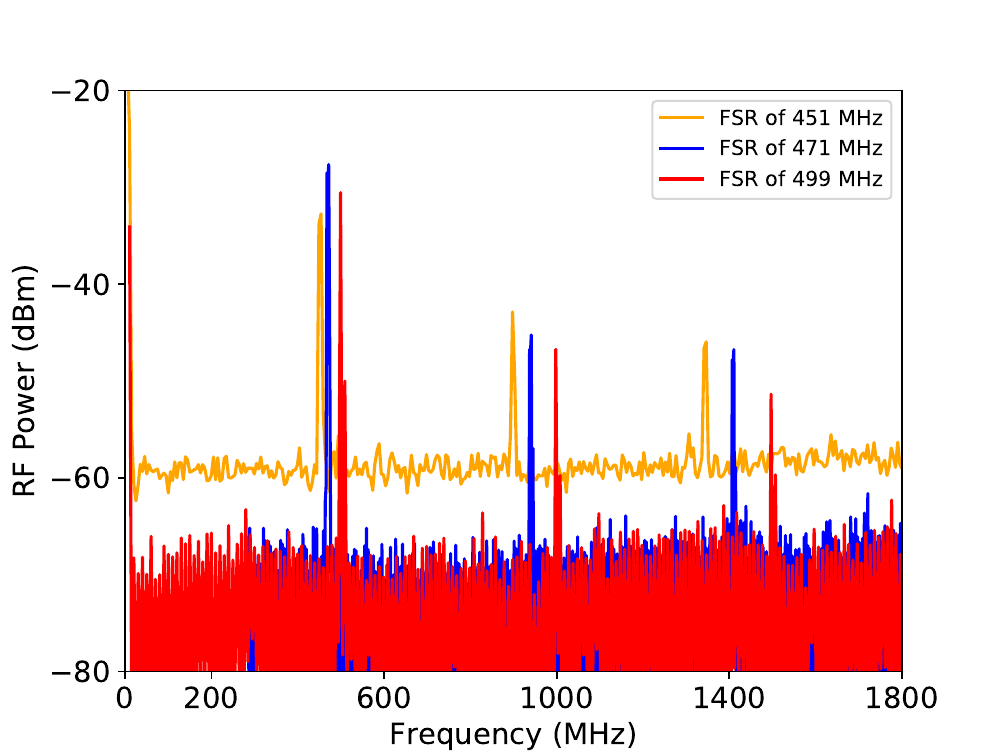}
    \end{subfigure}
      \caption{(a) Optical spectrum obtained at different tunings of the MRRs and the phase section (b) Corresponding RF spectra showing mode-locking at 451, 471, and 499 MHz. For mode-locking at 451 MHz, MRR$_1$ is tuned to 43.87 mW, MRR$_2$ to 151.25 mW, MRR$_3$ to 125.76 mW, and the phase section to 138.95 mW. For 471 MHz, the tunings are 39.55 mW, 153.92 mW, 108.80 mW, and 172.15 mW, respectively. For 499 MHz, the tunings are 43.87 mW, 151.25 mW, 98.29 mW, and 176.15 mW.}
    \label{fig:MLL}
\end{figure}


To address this question, we used detunings of the MRRs followed by optimizing the phase section as described above. Typically, one of the small MRRs is slightly detuned causing multimode operation before the phase section is tuned to restore mode-locking. As an example, to obtain mode-locking  starting from single frequency oscillation (MRR$_1$ tuned to 43.87 mW, MRR$_2$ to 151.25 mW, MRR$_3$ to 126.57 mW, and phase section tuned to 101.94 mW), the heater settings were modified to obtain mode locking with three different repetition rates (see Fig. \ref{fig:MLL}, which includes the heater settings for each repetition rate). Figure \ref{fig:MLL} shows that the optical spectra remain similar (due to limited resolution) while the RF spectra reveal mode-locking at various repetition rates: 451 MHz, 471 MHz and 496 MHz (selected as three typical cases with repetition rates less than 500 MHz). Also repetition rates higher than 500 MHz upto up to 650 MHz, were obtained with further detuning of the ring resonators, though higher repetition rates are not subject to investigation here.\\



\subsection{Hybrid mode-locking}
Passive, absorber-free mode-locking as described in sec. 2.3, is a simple and a robust method because it does not require any additional components such as electro-optic or acousto-optic modulators. However, passive mode-locking bears the drawback of the repetition rate being subject to drift and pulse jitter. Jitter is caused by intrinsic effects such as spontaneous emission combined into the laser modes in combination with gain-index coupling in semiconductor optical amplifier. Also external effects lead to instabilities such as acoustic perturbations and temperature drift that change the cavity length on various time scales. For instance, in our experiments we observe a small thermal drift of the repetition rate in the order of 20 MHz/min. To increase the stability of the repetition rate, we take a well-known hybrid mode-locking approach, where passive mode-locking is supported and stabilized with an external RF oscillator. If the frequency of the oscillator is chosen close enough to the passive repetition rate, within the so-called locking range, the optical roundtrip rate assumes the externally applied frequency \cite{diodelaserLR, LockingRangeAssymetry}. \\

\begin{figure}
    
    \begin{minipage}[]{0.9\textwidth}
    \centering
        \begin{subfigure}{0.5\textwidth}
            \centering
            \includegraphics[width=1\linewidth]{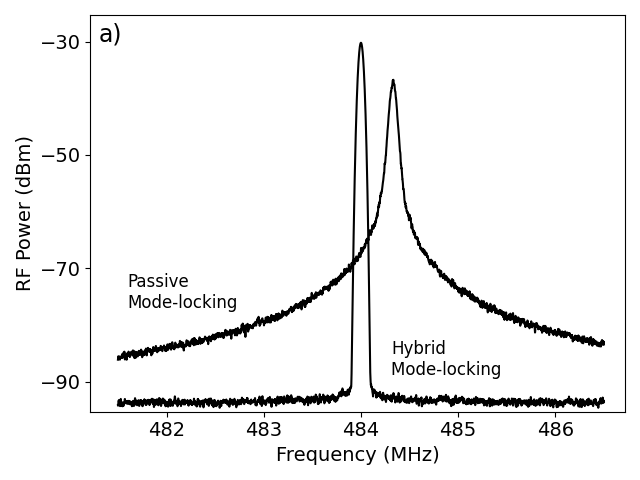}
           \end{subfigure}
    \end{minipage}
     \vspace{1em} 
    \begin{minipage}[t]{0.9\textwidth}
        \begin{subfigure}{0.5\textwidth}
            \centering
            \includegraphics[width=1\linewidth]{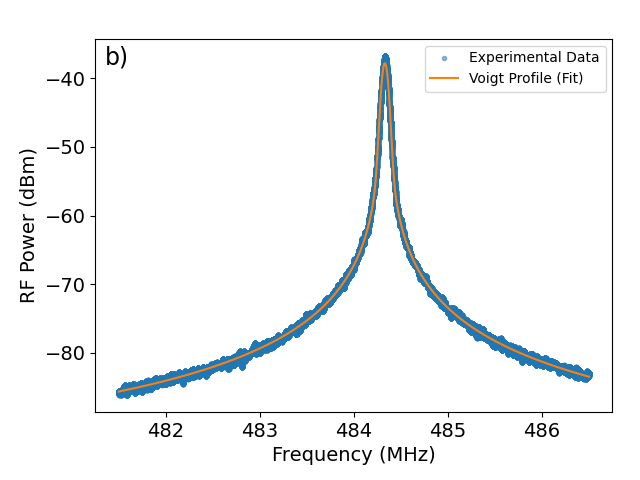}
            
        \end{subfigure}
        \begin{subfigure}{0.5\textwidth}
            \centering
            \includegraphics[width=1\linewidth]{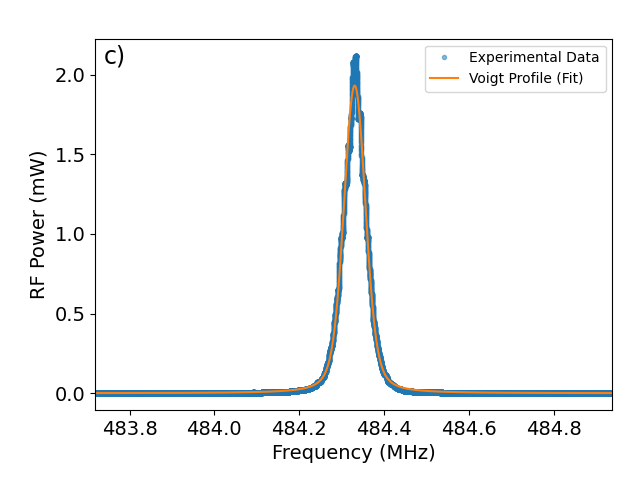}
          \end{subfigure}
    \end{minipage}
    
    \vspace{1em} 
   
    \begin{minipage}[t]{0.9\textwidth}
        \begin{subfigure}{0.5\textwidth}
            \centering
            \includegraphics[width=1\linewidth]{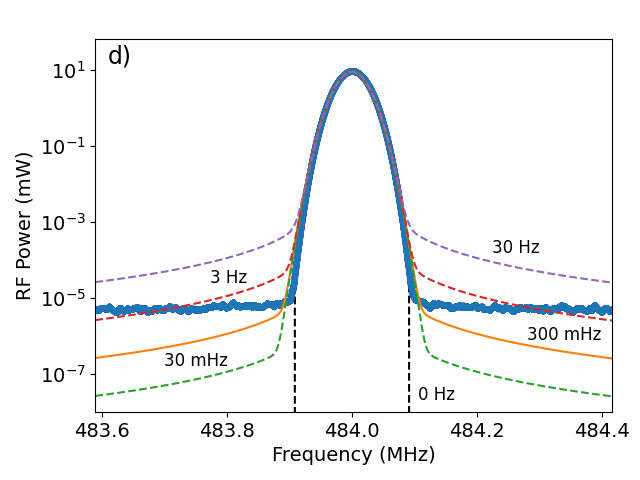}    
            \end{subfigure}
        \begin{subfigure}{0.5\textwidth}
            \centering
            \includegraphics[width=1\linewidth]{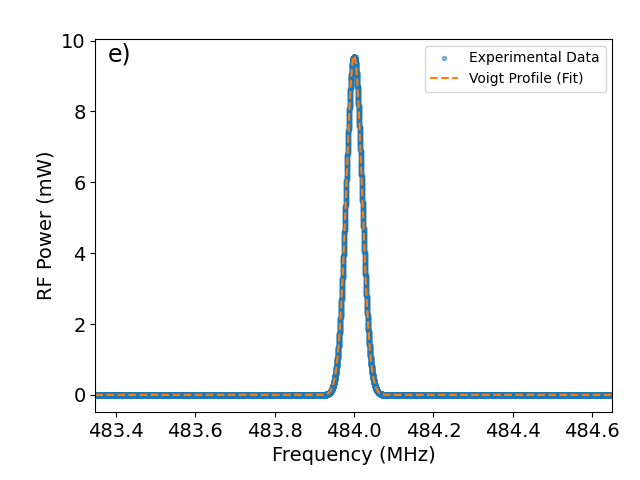}
    \end{subfigure}
\end{minipage}

\caption{a) RF spectra obtained with passive and hybrid mode-locking at repetition rates near 484 MHz. The experimental noise floor is at around -95 dB with the resolution and video bandwidth set to 47 kHz. b) Passive mode-locking: A logarithmic Voigt fit which gives high weight to the line wing data yields a Lorentzian component of 4.8 kHz (Gaussian linewidth 27.6 kHz with negligible errors). c) A Voigt fit to the same data on a linear scale puts the weight on the line centered data and thus yields different linewidths. Gaussian linewidth 21.4 kHz (Lorentzian linewidth 8.9 kHz again with negligible statistical errors). d) Hybrid mode-locking: The black dotted line presents a pure Gaussian profile with 19.9 kHz linewidth (obtained with a Voigt fit on a linear scale, refer e). The other dashed curves are Voigt profiles with a constant 19.9 kHz Gaussian component and various Lorentzian components between 30 mHz and 30 Hz. The solid orange line is a Voigt profile with 300 mHz Lorentzian component, denoting the upper limit beyond which the line wings would rise above the noise floor, which were inconsistent with the experimental data. e) RMS Voigt fit to the same data on a linear scale yielding a 19.9 kHz Gaussian component with near-zero Lorentzian part.}
\label{fig:LW comparision}
\end{figure}

For hybrid mode-locking, we add a sinusoidal modulation to the laser pump current using a frequency tunable RF generator (see setup Fig. \ref{fig:Setup}). Initially we applied a fixed modulation power of 1 mW (corresponding to 0.7 mA) and a DC pump current of 50 mA (as used in measurements for Fig. \ref{fig:PML}, \ref{fig:Multimode} and \ref{fig:MLL}). First the laser was set to passive mode-locking at 484.3 MHz with the RF oscillator turned off. Then the RF frequency was set to a similar value (484 MHz) and the RF current was applied, while monitoring the laser's RF spectrum. Figure \ref{fig:LW comparision} displays the RF spectra measured with passive and hybrid mode-locking. Their comparison shows that passive mode-locking entails a broad peak with wide line wings, whereas RF injection yields a stabilization of the repetition rate seen as reducing the wings to a level at or below the experimental noise floor. \\

For quantifying the linewidth reduction we performed a Voigt fit analysis with the results summarized in Fig. \ref{fig:LW comparision}b-e. For the case of passive mode-locking, we fitted a logarithmic Voigt profile (relative RF power in dB) to the measured RF spectrum, see Fig. \ref{fig:LW comparision}b. The RMS (root-mean-square) fit shows excellent agreement, yielding a Gaussian linewidth (21.4 kHz) comparable to the resolution, and a Lorentzian width of 8.9 kHz.  Fitting a Voigt profile to the same data, now on a linear vertical scale (Fig. \ref{fig:LW comparision}c), delivers somewhat different linewidth values, 27.6 kHz and 4.8 kHz, respectively. The reason for these differences is that the logarithmic fit is more precise for the Lorentzian linewing component as there most of the experimental data is located. The linear fit is more precise for the Gaussian component because it minimizes deviations of the highest-power values, which are found near line center. The deviations between the respective Gaussian and Lorentzian components show that the linewidth is not a pure Voigt profile. \\

For the case of hybrid mode-locking a logarithmic RMS Voigt fit did not provide reliable linewidth values, because the Lorentzian line wings were at or below the experimental noise floor. To determine an upper limit for the Lorentzian part, Fig. \ref{fig:LW comparision}d shows the measured RF spectrum with a series of logarithmic Voigt profiles. These have a fixed Gaussian width (19.9 kHz obtained with a linear Voigt fit) and a stepwise increased Lorentzian width. It can be seen that if the Lorentzian linewidth exceeds an upper limit of around 300 mHz, the Lorentzian wings rise above the noise level, which is inconsistent with the experimental data. Summarizing, we observed that mode-locking reduced the Lorentzian linewidth by at least three orders of magnitude, down to a detection limited upper value of 300 mHz. \\

\subsection{Locking Range}

In this section we present a measurement of the locking range, which is important for long-term stability of the repetition rate. If the locking range is bigger than the interval within which the passive repetition rate drifts, hybrid mode-locking might be maintained indefinitely, in spite of cavity drift. With passive mode-locking we observed that the repetition rates typically drifts within an interval smaller than 50 MHz with a speed below 20 MHz/min. \\


\begin{figure}[h!]
\centering\includegraphics[width=13cm]{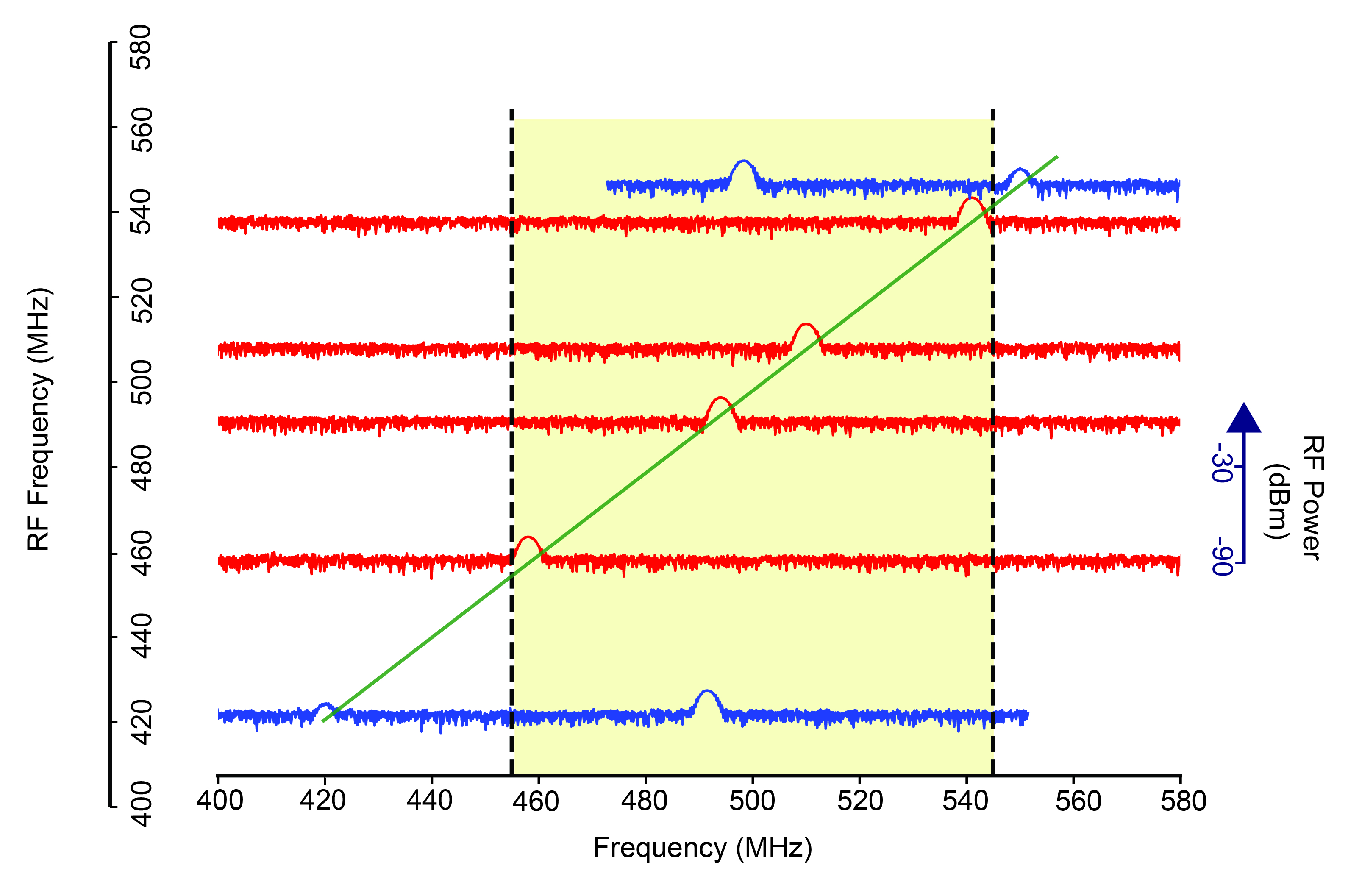}
\caption{RF spectra of the mode-locked laser recorded with 1 mW diode pump current modulation applied with various RF frequencies in the range between 420 and 550 MHz. The recorded  spectra are offset with the applied modulation frequency (left-hand side axis). The power of the recorded laser spectra is shown at the right-hand side axis (noise levels at - 90 dBm). The top and bottom spectra (blue traces) show two peaks, i.e., the laser is still passively mode-locked, as the modulation frequency lies outside the locking range. The spectra that display only a single RF peak are plotted in red. With these modulation frequencies, the laser's passive repetition rate is locked to the frequency of the external RF oscillator (hybrid mode-locking). The green line is a linear fit connecting the positions of tunable peaks. The range within which the repetition is locked to that of the external modulation frequency is 82 MHz wide (indicated in yellow).}
\label{fig:LockingRange}
\end{figure}


The experiments are carried out by incrementing the external modulation frequency across the passive repetition rate, while recording RF spectra. First, we measured the size of the locking range with a fixed RF modulation power of 1 mW. Figure \ref{fig:LockingRange} presents examples of recorded RF spectra, plotted with a vertical offset (left hand side vertical scale) equal to the  frequency selected at the RF oscillator. The right hand side scale shows the RF power of the laser spectra. Before applying RF modulation, the repetition rate of passive mode-locking was set to a value at around 484 MHz. For the lowest applied frequency of 420 MHz (blue trace at the bottom) the spectrum shows a modulation peak at 420 MHz. The second, larger, fixed- frequency peak at 484 MHz indicates that the passive mode-locking is still present. When tuning the external oscillator closer to the fixed-frequency peak, at approximately 38 MHz distance (i.e., with the external modulation set to 458 MHz, second trace from bottom), the passive mode-locking peak vanishes, leaving only a single RF peak in the spectrum. This proves that the external frequency has entered the locking range causing hybrid mode-locking at the frequency of the external oscillator. When tuning the external frequency right onto the passive mode-locking frequency (modulation frequency 490 MHz, third trace from bottom), hybrid mode-locking remains still distinguishable from passive mode-locking via strong linewidth narrowing as in Fig. \ref{fig:LW comparision}. At about 545 MHz, the passive mode-locking peak re-appears (near 500 MHz), in addition to RF modulation (uppermost trace). The frequency of hybrid mode-locking follows the modulation frequency as can be seen from the linear fit line that follows well the position of the tunable peaks. In these experiments, we find an RF locking range with a full width of approximately 82 MHz. \\ 


\begin{figure}[h!]
\centering\includegraphics[width=14cm]{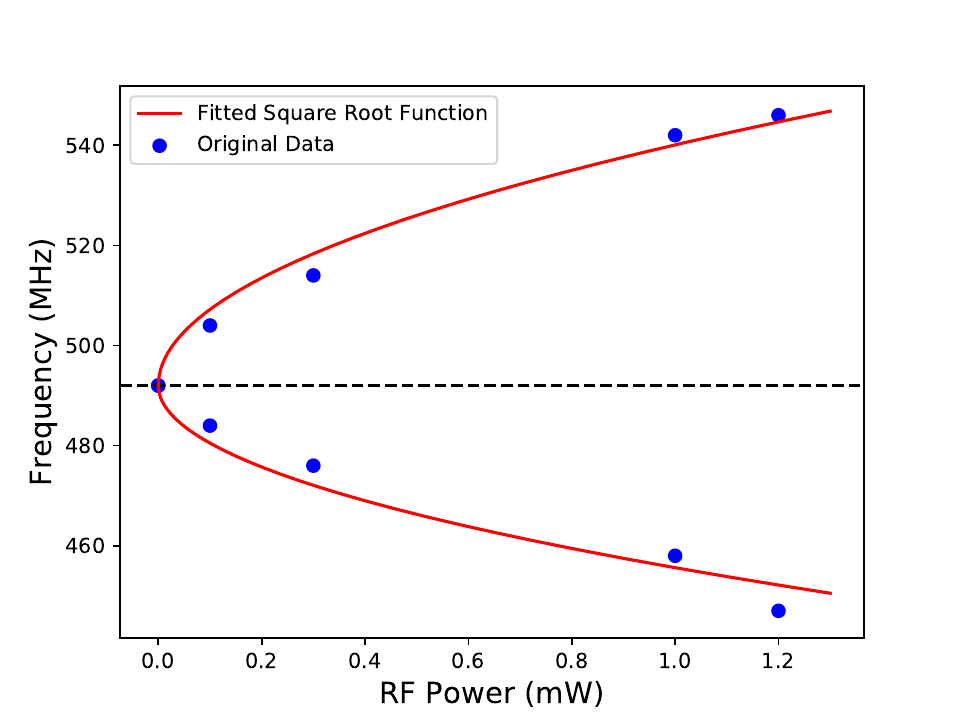}
\caption{Locking range (MHz) measured vs injected RF power. The black dotted line in the center shows the passive mode-locking frequency at 492 MHz. The curves are two square-root fit functions with different curvature, centered at the passive mode-locking frequency.}
\label{fig:RFpowerchangevsLockingrange}
\end{figure}

A second, well-know signature of injection locking is that the locking range should vary with the injected power following a square-root dependence \cite{diodelaserLR}. Specifically with diode lasers, the locking range is expected to increase asymmetrically with injected power which is due to gain-index coupling in semiconductor amplifiers \cite{LockingRangeAssymetry}. \\


To verify the presence of these properties and for a more quantitative characterization, we measured the width of the locking range as a function of the applied RF power. In the experiments, after establishing passive locking at around 492 MHz, the RF power was increased in steps from -10 dBm (0.1 mW) to 1 dBm (1.2 mW), and the edges of the locking range were determined as described in Fig. \ref{fig:LockingRange} (disappearance and reappearance of the passive repetition rate, observable with a typical experimental error of a few MHz). The results are summarized in Fig. \ref{fig:RFpowerchangevsLockingrange}. The horizontal black dashed line indicates the frequency of passive mode-locking. It can be seen that the lower edge of the locking range decreases in frequency (45 MHz) while the upper edge increases in value (54 MHz). It can also be seen that the edges of the locking range vary asymmetrically, following approximately two square-root fit functions plotted for comparison. For hybrid mode-locking of diode lasers with semiconductor saturable absorbers it has been shown \cite{diodelaserLR} that the gain-index coupling parameter (Henry's linewidth enhancement factor) of both the gain section and the saturable absorber contribute to an asymmetric RF locking range. As in our experiments no saturable absorber is present, the observed asymmetry can thus be addressed solely to gain-index coupling in the semiconductor amplifier. \\    

\subsection{Relation between pump current and heater settings}

The experiments described so far used a fixed DC pump current of 50 mA, which is moderately above the threshold current (27 mA). Theoretical modelling has predicted a high robustness of the observed Fourier domain mode-locking upon increasing the pump current\cite{GioanniniHighPumpCurrent}. For a laser of similar design, though with about ten-times higher repetition rate, it was predicted that, beyond about five-times the threshold, mode-locking is maintained with a constant setting of the phase section when further increasing the pump current. This prediction is, however, based on a simplified description that neglects spatial gain variations or thermally induced phase shifts in the gain section, which might terminate mode-locking when increasing the pump current.\\


\begin{figure}[ht!]
\centering\includegraphics[width=14cm]{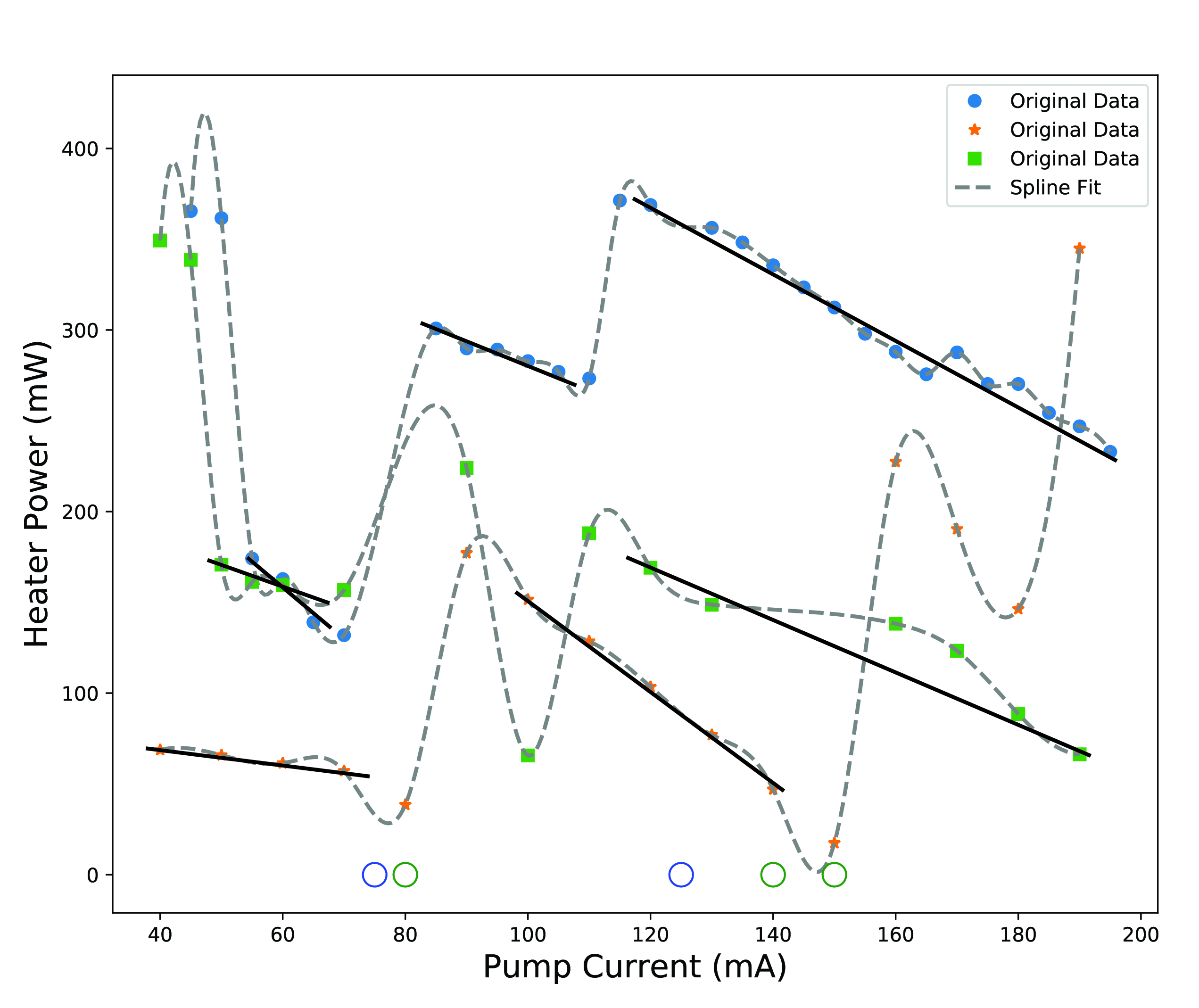}
\caption{Measured heater settings of the phase section required for stable mode-locking when increasing the diode pump current. Three different data sets are shown (blue, green and red symbols) grouped with dashed curves (spline fit), as obtained with three independent measurement runs. Open circles represent pump currents where mode-locking could not be found. It can also be seen that the absolute values for the required heater currents are not well reproducible between the different datasets (recorded at different days). We address this to experimental uncertainties, specifically to a residual thermal drift of the ring resonators. The straight lines are linear fits on subsets of data, showing an average slope of about -1.7 mW/mA.}
\label{fig:powerheatervspumpcurrent}
\end{figure}


To examine these predictions, we investigate whether mode-locking can be maintained toward increasing pump current, which is also of interest for providing frequency combs with increased output power. The experiment was carried out as follows. First stable mode-locking was established as described in Sect. 2.3, then the pump current is increased in steps. With each increase, we observe that the mode-locking became either unstable or terminated, which is different than predicted \cite{Mariangela}. However, the mode-locking became re-established by re-adjusting the phase section heater current, generally to some lower value. Figure \ref{fig:powerheatervspumpcurrent} shows the measured heater power required to re-establish mode-locking vs step-wise increasing the diode laser pump current from 50 mA to 190 mA. The figure presents three different data sets, obtained with three independent measurement runs with same settings spanning approximately the same range of pump currents. Straight lines with an average slope of -1.7 mW/mA were drawn if more than two data points lie in proximity before any steep dip or rise is seen in a measurement run. The data show that the heater current follows an approximately linear relation for most of the data points.\\

Our interpretation of the observed linear variation is that increasing the diode pump current does not only increase the gain and output of the laser, but it also increases the temperature of the InP p-n junction, roughly in proportion with the current. Via the thermo-optic coefficient of the semiconductor material, this increases the laser cavity roundtrip phase. To maintain mode-locking, the increased phase needs to be compensated by reducing the phase shift in the Si$_3$N$_4$ phase section. We note that there the phase shift is approximately proportional to the electric power, i.e., to the square of the heater current. We find that the negative sign of the slope in Fig. \ref{fig:powerheatervspumpcurrent} and its value are in reasonable agreement with typically expected temperature changes that are on the order of 20 deg in the InP p-n junction \cite{TempvsCurrentInP} and in the order of 200 deg in the Si$_3$N$_4$ phase section \cite{TempVsCurrentSiN}. For this we have considered the ratio of the respective thermo-optic coefficients \cite{ThermoOpticINP, ThermoOpticSiN} and the lengths of the gain section (700 µm) and the phase section (1 mm). \\

\section{Summary and Conclusion}
Using a hybrid integrated diode laser, we demonstrate absorber-free passive mode-locking at sub-GHz repetition rates as low as 451 MHz. The laser comprises an InP based semiconductor optical amplifier and a Si$_3$N$_4$ feedback circuit. A low repetition rate is provided via a long effective laser cavity, obtained via a feedback circuit with three double-passed microring resonators as spectral Vernier filter. Passive mode-locking is achieved by slightly increasing the cavity length via an integrated phase section, when the laser is initially in single-frequency operation. The repetition rate of the laser is tunable across a wide range from about 450 to 650 MHz. This is achieved by tuning the microring resonators slightly off mutual resonance, as this reduces the number of roundtrips and thereby shortens the effective laser cavity length. To stabilize the repetition rate of the laser, we add an RF current to the DC diode pump current, which induced hybrid mode-locking at the external RF frequency. The corresponding stabilization of the repetition rate shows as reduction of the Lorentzian linewidth component from 8.9 kHz towards to a detection-limited value around 300 mHz. This compares well with the sub-Hz 10 dB linewidth observed earlier with pulsed mode-locking at higher repetition rate \cite{dense-combs}. We measure a locking range of about 80 MHz with 1 mW RF power. The locking range increases with the injected RF power following an asymmetric square root dependence as expected. Absorber-free, passive mode-locking was observed up to the maximum specified diode pump current, by re-adjusting the cavity via the phase section. Regarding future developments, earlier numerical calculations have indicated that a further lowering of the repetition rate should be possible \cite{JesseThesis} using further length extension. Following this route promises to bring the repetition rate near that of widely used bulk lasers.

      
\begin{backmatter}
\bmsection{Acknowledgments}
The authors gratefully acknowledge Yvan Klaver, Lisa Winkler and Redlef Braamhaar for fruitful discussions on mode-locking and vital technical support throughout this work.

\bmsection{Funding}
This project is partly funded by the Netherlands Organisation for Scientific Research (NWO) under the project title: "Ultra-narrowband lasers on a chip" and partly by the European Union’s Horizon 2020 research and innovation program under grant agreement No. 780502 (3PEAT).

\bmsection{Disclosure}
The authors declare no conflicts of interest.

 \bibliography{Manuscript}





\end{backmatter}
\end{document}